\documentclass[a4paper,10pt]{article}
	\usepackage{enumerate}
	\usepackage{color}
	\usepackage[utf8]{inputenc} 
	\usepackage[english]{babel}
	\usepackage[T1]{fontenc}
	\usepackage{graphicx}
	\usepackage{empheq} \usepackage{amsfonts,amssymb,amsmath,latexsym,amsthm}
	\usepackage{textcomp}
	\usepackage[pdftex]{hyperref}
	\usepackage{geometry}
	\usepackage{tikz}
 \usepackage{bm}

	\author{L. G. Martins$^1$,
 M. V. Flamarion$^{2,*}$,  R. Ribeiro-Jr$^3$}
 
	\title{Flow Structures Beneath Stationary Waves with Constant Vorticity Over Variable Topography}

	\date{}
	
\begin{document}
	\maketitle
	\begin{center}

		{\footnotesize $^1$ UFPR/Federal University of Paraná \\
	luiz.martins1@ufpr.br }

		\vspace{0.3cm}

		{\footnotesize $^{2}$
  { Departamento Ciencias–Sección Matemáticas, Pontificia Universidad Católica del Perú, Av. Universitaria 1801, San Miguel 15088, Lima, Peru \\
	corresponding author$^{*}$: mvellosoflamarionvasconcellos@pucp.edu.pe}

		\vspace{0.3cm}

		{\footnotesize $^3$
  UFPR/Federal University of Paran\'a,  Departamento de Matem\'atica, Centro Polit\'ecnico, Jardim das Am\'ericas, Caixa Postal 19081, Curitiba, PR, 81531-980, Brazil \\ robertoribeiro@ufpr.br}
	}
	

	\end{center}
	
	\begin{abstract} 
        \noindent 
The flow structures beneath waves have received significant attention from both theoretical and numerical perspectives. Most studies on this topic assume a flat bottom, leading to  questions about the effects of variable bottom topography. To address this gap, we investigate the flow structures beneath stationary waves with constant vorticity, considering the influence of variable topography. Specifically, we numerically analyze the role of vorticity in the emergence of stagnation points and the pressure distribution within the fluid in two bottom topography scenarios: a bump and a hole.
Our numerical approach is based on a variation of the classical Dyachenko, Zakharov, and Kuznetsov conformal mapping technique for free-boundary water wave problems.
Our results reveal the existence of  saddle points beneath wave crests and center beneath depression solitary waves. Additionally, we observe that the pressure can exhibit distinctive features, such as a global minimum on the bottom boundary -- behavior that is markedly different from the usual  flat-bottom case.

	\end{abstract}

\section{Introduction}

In 1996, Zakharov collaborated on two seminal studies that introduced a novel conformal mapping technique for addressing water wave problems. One paper, co-authored with Dyachenko and Kuznetsov, established the foundational ideas behind this approach for fluid domains of finite depth \cite{Dyachenko et al:1996b}. The other paper, co-authored with Dyachenko, Kuznetsov, and Spector, applied these concepts to the case of infinite depth \cite{Dyachenko et al:1996a}. These works presented a conformal mapping technique that encapsulates the geometric complexity of the moving free surface into a time-dependent conformal map. This transformation simplifies the moving boundary problem by reducing it to static domains, such as a strip for the finite-depth regime and a half-plane for the infinite-depth regime. By employing the Fourier transform, their method proved highly effective for numerical simulations. For simplicity, we refer to the conformal map developed by the authors previously mentioned as the DKSZ Conformal Mapping, regardless of the specific depth regime.

The ideas employed in the DKSZ mapping laid the groundwork for the development of pseudospectral numerical methods for free-surface hydrodynamic problems that continue to this day. These methods have enabled significant advancements in understanding of free-surface wave shapes, integral properties such as wave energy and momentum, and the flow structures beneath the waves. They have been applied to various contexts, including gravity waves \cite{ChoiCamassa:1999,Dyachenko:2014, Dyachenko:2021,Li:2004, Lushnikov:2021, Zakharov:2020},
capillary-gravity waves \cite{Shelton:2023, Milewski:2010}, hydroelastic waves \cite{Milewski:2011,Tao:2019, Wang:2011}, rotational waves in   constant vorticity  flows \cite{Choi:2009,Castro:2023,DyachenkoHur2:2019,Ribeiro:2017}, electrohydrodynamic waves \cite{Tao:2023,Tao:2018,Kochurin:20234,Doak:2022,Zubarev:2025,Flamarion:2024},  wave dynamics over a variable bottom topography \cite{Flamarion:2019,Viotti:2014,Flamarion:2021} turbulence \cite{Dyachenko:2004, Kachulin:2020, Kochurin:2023, Kochurin:2020, Kochurin:2020a, Kochurin:2025, Kochurin:2025a, Onorato:2002} and freak waves \cite{Dyachenko:2014, Dyachenko:2005, Zakharov:2002,Zakharov:2006, Zakharov:2013}. We refer the reader to Nachbin \cite{Nachbin:2023} for an alternative approach to addressing water wave problems in flows over variable bottom topography.

Among the studies that leverage the DKSZ mapping, we focus on the contributions of Flamarion {\it et al.} \cite{Flamarion:2019} and Ribeiro-Jr {\it et al.}  \cite{Ribeiro:2017}. The numerical method developed by Flamarion {\it et al.}  enables the analysis of wave dynamics over variable topography, while Ribeiro-Jr {\it et al.}'s approach facilitates the study of flow structures beneath waves on a fluid with constant vorticity over flat bottoms. In this paper, we integrate these two methods to investigate the flow structure beneath stationary nonlinear surface waves influenced by bottom topography on a fluid with constant vorticity.

Our study specifically examines the emergence of stagnation points and pressure anomalies. Stagnation points are locations where fluid particles exhibit zero velocity in the wave’s moving frame, corresponding to stationary particles in the case of a stationary wave. Pressure anomalies, in turn, is a term used to describe flows where the minimum pressure is located within the bulk of the fluid or when the pressure does not strictly increase with depth along vertical lines.

It is well-known that the flow beneath a rotational water wave with constant vorticity is primarily characterized by the appearance of stagnation points and the emergence of pressure anomalies. These phenomena have been demonstrated numerically \cite{Choi:2003, Ribeiro:2017,Teles:1988,VasaneOliveras:2014,Castro:2023}, derived asymptotically  \cite{AliKalish:2013,Johson:1986}
and rigorously proven \cite{EhrnstromVillari:1888,Kozlov:2020,Wahlen:2009}.
Constant vorticity also influences the shape of the free surface significantly; for instance, the wave profile can become rounder and may even form exotic overhanging waves \cite{DyachenkoHur2:2019,Broeck:1994,Broeck:1996,DyachenkoHur:2019,ConstantinStrauss:2016,HurWheeler:2022}. However, all these results are based on the assumption of flow over a flat bottom.

Physically, models for waves with constant vorticity provide realistic approximations for scenarios such as water waves that are long compared to the water depth or short relative to the length scale of the vorticity distribution \cite{Teles:1988}. Despite their physical relevance, the theoretical and numerical literature on the flow structure beneath waves with constant vorticity  over a variable bottom topography is scarce. To the best of our knowledge, no studies have explored this topic within the framework of the full Euler equations. Existing research has primarily focused on the context of the linearized Euler equations \cite{Flamarion:2020}. 

This study aims to bridge this gap and advance the understanding of flow structures beneath waves influenced by constant vorticity in flows over  variable bottom topography. Our numerical simulations identify parameter regimes where unique flow behaviors emerge, including the presence of saddle points beneath wave crests and pressure at the bottom being lower than at the free surface—phenomena absent in flows over flat bottoms. These findings highlight the intricate interplay between wave dynamics, vorticity, and variable topography.

The present article is structured as follows. The mathematical formulation and numerical scheme are described in section 2 and 3. The findings are presented and discussed in section 4. Lastly, the final considerations are made in section 5.

\section{Mathematical Formulation}

\subsection{Euler Equations}

Consider an inviscid, and incompressible fluid with a constant density $\rho$ in a two dimensional space $(x,y)$. We establish a Cartesian coordinate system $(x,y)$ with the gravity $g$ pointing in the negative $y-$direction and $y = 0$ being the undisturbed free surface. The bottom boundary is localized at $y = -h_0 + h(x)$ and the upper boundary is free to move and denoted by $\zeta(x,t)$. We suppose the bottom obstacle $h$ and the pressure on the free surface $P$ are moving horizontally with constant speed $U$.

We are interested in studying flows with a constant vorticity. With this in mind, we write the velocity field as $(u,v) = \nabla \phi + (ay,0)$,  where $\phi$ is the velocity potential of an irrotational perturbation of the shear flow and  $a$ is the constant vorticity. By making a convenient change of variables, we have a scenario where both the topography and the surface pressure are stationary and there is a linear current that varies vertically.

As presented by Flamarion  {\it et al.}   \cite{Flamarion:2020}, the dimensionless velocity field read as
\begin{equation} \label{eq:shearflow}
    (u,v) = \nabla \phi + (\Omega y + F,0),
\end{equation}
and the dimensionless Euler equations are given by
\begin{empheq}[left=\empheqlbrace]{align}
    &\Delta \phi = 0, &\text{in} \ -1 + h(x) < y < \zeta(x,t) \label{laplace} \\ 
    &(F - \Omega)h_x + \Omega hh_x + \phi_x h_x = \phi_y, &\text{at} \ y = -1 + h(x) \label{noslip}  \\
   &\zeta_t + (F+\Omega\zeta + \phi_x)\zeta_x - \phi_y = 0, &\text{at} \ y = \zeta(x,t) \label{kinem} \\
   &\phi_t + \dfrac{1}{2}\left(\phi_x^2 + \phi_y^2 \right) + (F+\Omega\zeta)\phi_x + \zeta - \Omega\psi = -P(x), &\text{at} \ y = \zeta(x,t) \label{bernoulli} 
\end{empheq}
where $\Omega = ah_0/(gh_0)^{1/2}$ and $F = U/(gh_0)^{1/2}$ are, respectively, the dimensioneless vorticity and the Froude number. In the following subsections, we describe the conformal mapping technique used to solve the Euler equations, as well as our approach to compute the velocity field and pressure within the bulk of the fluid. The numerical conformal mapping technique combines the methodologies presented in the works of Flamarion  {\it et al.}   \cite{Flamarion:2019} and Ribeiro-Jr {\it et al.} \cite{Ribeiro:2017}. For further details, we refer readers to these references.

\subsection{Conformal mapping}

We assume that 
\begin{equation*}
    |\zeta(x,t)| \rightarrow 0, |\phi(x,t)|\rightarrow 0, |\psi(x,t)|\rightarrow 0, |h(x)| \rightarrow 0 \quad \text{and} \quad |P(x)| \rightarrow 0
\end{equation*}
    as $|x| \rightarrow +\infty$, so we can truncate the unbounded domain to a bounded one, and approximate the boundary conditions by periodic conditions. Now, to solve Euler equations \eqref{laplace}-\eqref{bernoulli}, we use a conformal mapping technique.

We introduce in the canonical domain a coordinate system $(\xi,\eta)$. The conformal map is given by
\begin{equation*}
    z(\xi,\eta,t) = x(\xi,\eta,t) + iy(\xi,\eta,t)
\end{equation*}
and it satisfies the following conditions
\begin{equation*}
    y(\xi,0,t) = \zeta(x(\xi,0,t),t) \quad \text{and} \quad y(\xi,-D,t) = -1 + H(\xi,t),
\end{equation*}
where $H(\xi,t) = h(x(\xi,-D,t))$. In other words, we impose that the strip $\eta = 0$ is mapped to the free surface $y = \zeta(x,t)$ and that the strip $\zeta = -D$ is mapped onto the topography profile.

Lets denote by $\mathbf{X}(\xi,t)$ and $\mathbf{Y}(\xi,t)$ the horizontal and vertical component of the conformal map in the free surface, and by $X_b(\xi,t)$ the horizontal component of the conformal map at $\eta = -D$. We note that the conformal map preserves the Laplacian equation. Therefore, we obtain the following system
\begin{equation*}
    \begin{cases}
        y_{\eta\eta} + y_{\xi\xi} = 0, &\text{in} \ -D < \eta < 0,  \\
    y(\xi,0,t) = \mathbf{Y}(\xi,t), \\
    y(\xi,-D,t) = -1 + H(\xi,t), 
    \end{cases}
\end{equation*}
whose solution is
\begin{align}
     y(\xi,\eta,t) &= \mathcal{F}_{k \neq 0}^{-1} \left[ \dfrac{\sinh(k(D+\eta))}{\sinh(kD)}\hat{\mathbf{Y}}(k,t)\right] -\mathcal{F}_{k \neq 0}^{-1} \left[ \dfrac{\sinh(k\eta)\coth(kD)}{\cosh(kD)} \hat{H}(k,t)\right] \notag \\
     &\hspace{.8cm}+ \left(\dfrac{1-\hat{H}(0,t)+\hat{\mathbf{Y}}(0)}{D}\right)\eta + \hat{\mathbf{Y}}(0,t), \label{y-cd}
\end{align}
where $\mathcal{F}_{k\neq 0}^{-1}$ denotes the inverse of Fourier transform without the zero mode.

It follows from the Cauchy-Riemann equations and some basic computations that
\begin{align}
    x(\xi,\eta,t) &=  \mathcal{F}_{k \neq 0}^{-1} \left[ i\left(\dfrac{\cosh(k\eta)\coth(kD)}{\cosh(kD)}\right)\hat{H}(k,t)\right] - \mathcal{F}_{k \neq 0}^{-1} \left[i\left(\dfrac{\cosh(k(D+\eta))}{\sinh(kD)}\right)\hat{\mathbf{Y}}(k,t)\right] \notag \\
    &\hspace{.8cm} + \left(\dfrac{1-\hat{H}(0,t) + \hat{\mathbf{Y}}(0,t)}{D}\right)\xi.   \label{x}
\end{align}
From the expression \eqref{x}, we can derive
\begin{align}
    X_b(\xi,t) &= \mathcal{F}_{k \neq 0}^{-1}[i\tanh(kD)\hat{H}(k,t)] + \mathcal{F}_{k \neq 0}^{-1}\left[ i\coth(kD)\left(\dfrac{\hat{H}(k,t)}{\cosh^2(kD)} - \dfrac{\hat{\mathbf{Y}}(k,t)}{\cosh(kD)} \right) \right] \notag  \\
    &\hspace{.8cm} + \left(\dfrac{1-\hat{H}(0,t) + \hat{\mathbf{Y}}(0,t)}{D}\right)\xi.   \label{xbottom}
\end{align}
We impose that the physical domain has the same length as the canonical domain. Hence, we choose $D$ as
\begin{equation*}
  D = D(t) = 1-\hat{H}(0,t) + \hat{\mathbf{Y}}(0,t).
\end{equation*}

Denote by $\overline{\phi}$ the potential velocity and $\overline{\psi}$ its conjugate in the canonical domain. We have that both satisfy the Laplace equation with the following boundary conditions:
\begin{equation*}
    \begin{cases}
        \overline{\phi}_{\eta \eta} + \overline{\phi}_{\xi \xi} = 0, \ \text{in} -D < \eta < 0, \\
        \overline{\phi}(\xi,0,t) = \mathbf{\Phi}(\xi,t), \\
        \overline{\phi}_\eta(\xi,-D,t) = (F - \Omega)H_\xi + \Omega HH_\xi, 
    \end{cases} \quad \text{and} \quad  \begin{cases}
        \overline{\psi}_{\eta \eta} + \overline{\psi}_{\xi \xi} = 0, \ \text{in} -D < \eta < 0, \\
        \overline{\psi}(\xi,0,t) = \mathbf{\Psi}(\xi,t), \\
        \overline{\psi}(\xi,-D,t) = -(F-\Omega)H(\xi,t) - \dfrac{\Omega}{2}H^2(\xi,t) + Q(t), 
    \end{cases}
\end{equation*}
where  Q(t) will be determined a posteriori.

Solving these problems leads to
\begin{empheq}[left = \empheqlbrace]{align}
   \overline{\phi}(\xi,\eta,t) &=  \mathcal{F}^{-1} \left[\dfrac{\cosh(k(D+\eta))}{\cosh(kD)}\hat{\mathbf{\Phi}}(k,t) + \dfrac{\sinh(k\eta)}{\cosh(kD)}\left(i(F-\Omega)\hat{H}(k,t) + i\dfrac{\Omega}{2}\widehat{H^2}(k,t)\right)\right], \notag \\[4mm]
   \overline{\psi}(\xi,\eta,t) &= \mathcal{F}_{k \neq 0}^{-1}\left[\left(\dfrac{(F-\Omega)\hat{H}(k,t) + \frac{\Omega}{2}\widehat{H^2}(k,t)}{\cosh(kD)} + \hat{\mathbf{\Psi}}(k,t) \right)\dfrac{\sinh(k(D+\eta))}{\sinh(kD)}\right] \notag  \\
   &\hspace{.5cm}- \mathcal{F}_{k \neq 0}^{-1} \left[\left((F-\Omega)\hat{H}(k,t) + \dfrac{\Omega}{2}\widehat{H^2}(k,t)\right)\dfrac{\cosh(k\eta)}{\cosh(kD)}\right]  \notag \\
   &\hspace{.5cm}+ \left(\dfrac{\hat{\Psi}(0,t) + (F-\Omega)\hat{H}(0,t) + \frac{\Omega}{2}\widehat{H^2}(0,t) - Q}{D}\right)\eta + \hat{\mathbf{\Psi}}(0,t). \label{psi-cd}
\end{empheq}

From the Cauchy-Riemann equation 
\[
\overline{\phi}_\eta = -\overline{\psi}_\xi,
\]
we derive the following relation between \(\mathbf{\Phi}\) and \(\mathbf{\Psi}\):
\begin{align}
    \mathbf{\Phi}_\xi(\xi) &= -\mathcal{F}_{k \neq 0}^{-1} \big[i\coth(kD)\hat{\mathbf{\Psi}}_\xi(k,t)\big] 
    + \mathcal{F}_{k \neq 0}^{-1} \Bigg[ k \frac{\coth(kD)}{\cosh(kD)} \Big( (F - \Omega)\hat{H}(k,t) + \frac{\Omega}{2}\widehat{H^2}(k,t) \Big) \Bigg].
    \label{phi-psi}
\end{align}
On the other hand, the Cauchy-Riemann equation 
\[
\overline{\phi}_\xi = \overline{\psi}_\eta,
\]
leads to the expression
\begin{align}
    \mathbf{\Phi}_\xi(\xi, t) &= -\mathcal{F}_{k \neq 0}^{-1} \big[i\coth(kD)\hat{\mathbf{\Psi}}_\xi(k,t)\big] 
    + \mathcal{F}_{k \neq 0}^{-1} \Bigg[ k \frac{\coth(kD)}{\cosh(kD)} \Big( (F - \Omega)\hat{H}(k,t) + \frac{\Omega}{2}\widehat{H^2}(k,t) \Big) \Bigg] \nonumber \\
    &\quad + \frac{\hat{\mathbf{\Psi}}(0,t) + (F - \Omega)\hat{H}(0,t) + \frac{\Omega}{2}\widehat{H^2}(0,t) - Q}{D}.
    \label{phi-psi3}
\end{align}
Since equations \eqref{phi-psi} and \eqref{phi-psi3} must be equal, we conclude that the constant \(Q\) satisfies the relation:
\[
Q = \hat{\mathbf{\Psi}}(0,t) + (F - \Omega)\hat{H}(0,t) + \frac{\Omega}{2}\widehat{H^2}(0,t).
\]

Next, we write the boundary conditions on the free-surface  \eqref{kinem}-\eqref{bernoulli} in the canonical domain. After some manipulations, we derive the following evolution system
\begin{empheq}[left=\empheqlbrace]{align}
        &\mathbf{Y}_t = \mathcal{C}[\Theta(\xi,t)]\mathbf{Y}_\xi - \Theta(\xi,t)\mathbf{X}_\xi, \label{kc-cd} \\ 
        &\mathbf{\Phi}_t = - \mathbf{Y} + \Omega \mathbf{\Psi} +  \mathcal{C}[\Theta(\xi,t)]\mathbf{\Phi}_\xi - \dfrac{1}{2J}\left(\mathbf{\Phi}_\xi^2 - \mathbf{\Psi}_\xi^2 \right) - \dfrac{1}{J}(F+\Omega \mathbf{Y})\mathbf{\Phi}_\xi \mathbf{X}_\xi - P(\mathbf{X})
    \label{bc-cd}
\end{empheq}
where $J = \mathbf{X}_\xi^2 + \mathbf{Y}_\xi^2$ is the Jacobian on the free surface, $\Theta(\xi,t) = \dfrac{(F+\Omega \mathbf{Y})\mathbf{Y}_\xi + \mathbf{\Psi}_\xi}{J}$, and $ \mathcal{C}$ is the Hilbert operator which is defined as 
\begin{equation*}
    \mathcal{C}[f(\xi)] = \mathcal{F}_{k \neq 0}^{-1}[i\coth(kD)\mathcal{F}[f(\xi)]] + \lim_{k \rightarrow 0} i\coth(kD)\mathcal{F}[f(\xi)].
\end{equation*}

Gathering equations \eqref{x}, \eqref{phi-psi}-\eqref{bc-cd}, we obtain the Euler equations in the canonical domain
\begin{empheq}[left=\empheqlbrace]{align}
         &\mathbf{X}_\xi(\xi,t) =  1 +  \mathcal{F}_{k \neq 0}^{-1}\left[k\coth(kD)\hat{\mathbf{Y}}(k,t)\right] - \mathcal{F}_{k \neq 0}^{-1}  \left[k\dfrac{\coth(kD)}{\cosh(kD)}\hat{H}(k,t)\right], \label{xbottom-cd} \\[3mm]
        &\mathbf{\Phi}_\xi(\xi,t) =  -\mathcal{F}_{k \neq 0}^{-1}[i\coth(kD)\hat{\mathbf{\Psi}}_\xi(k,t)] + \mathcal{F}_{k \neq 0}^{-1} \left[ k\dfrac{\coth(kD)}{\cosh(kD)}\left( (F-\Omega)\hat{H}(k,t) + \dfrac{\Omega}{2}\widehat{H^2}(k,t) \right) \right], \\[3mm]
         &\mathbf{Y}_t = \mathcal{C}[\Theta(\xi,t)]\mathbf{Y}_\xi - \Theta(\xi,t)X_\xi, \\ 
        &\mathbf{\Phi}_t = - \mathbf{Y} + \Omega \mathbf{\Psi} +  \mathcal{C}[\Theta(\xi,t)]\mathbf{\Phi}_\xi - \dfrac{1}{2J}\left(\mathbf{\Phi}_\xi^2 - \mathbf{\Psi}_\xi^2 \right) - \dfrac{1}{J}(F+\Omega Y)\mathbf{\Phi}_\xi \mathbf{X}_\xi - P(\mathbf{X}).
    \label{eulercanonical}
\end{empheq}
We notice that in order to compute $H(\xi,t)$ we need to evaluate $h(X_b(\xi,t))$. However, as we can see in the equation \eqref{xbottom}, $X_b$ depends on $H$, so we have a implicit relation. So, we have to include the following equations
\begin{empheq}[left=\empheqlbrace]{align}
        &H(\xi,t) = h(X_b(\xi,t)), \label{add-eq1} \\
        &X_b(\xi,t) = \xi + \mathcal{F}_{k \neq 0}^{-1}[i\tanh(kD)\hat{H}(k,t)] + \mathcal{F}_{k\neq 0}^{-1}\left[i\coth(kD)\left(\dfrac{\hat{H}(k,t)}{\cosh^2(kD)}-\dfrac{\hat{\mathbf{Y}}(k,t)}{\cosh(kD)}\right)\right].
    \label{add-eq2}
\end{empheq}

In this work, we are interested in stationary solutions. So, we set all the partial derivatives in time to be zero. By carrying some computation, the system os equations \eqref{xbottom-cd}-\eqref{add-eq2} is written as
\begin{empheq}[left=\empheqlbrace]{align}
     &H(\xi) = h(X_b(\xi)), \label{st-eq1} \\
    &X_b(\xi) = \xi + \mathcal{F}_{k \neq 0}^{-1}[i\tanh(kD)\hat{H}(k)] + \mathcal{F}_{k\neq 0}^{-1}\left[i\coth(kD)\left(\dfrac{\hat{H}(k)}{\cosh^2(kD)}-\dfrac{\hat{Y}(k)}{\cosh(kD)}\right)\right], \\
     &\mathbf{X}_\xi(\xi) =  1 +  \mathcal{F}_{k \neq 0}^{-1}\left[k\coth(kD)\hat{\mathbf{Y}}(k)\right] - \mathcal{F}_{k \neq 0}^{-1}  \left[k\dfrac{\coth(kD)}{\cosh(kD)}\hat{H}(k)\right], \\[3mm]
     &\mathbf{\Psi}_\xi(\xi) = - (F+\Omega \mathbf{Y}(\xi))\mathbf{Y}_\xi(\xi), \\
          &\mathbf{\Phi}_\xi(\xi) =  -\mathcal{F}_{k \neq 0}^{-1}[i\coth(kD)\hat{\mathbf{\Psi}}_\xi(k)] + \mathcal{F}_{k \neq 0}^{-1} \left[ k\dfrac{\coth(kD)}{\cosh(kD)}\left( (F-\Omega)\hat{H}(k) + \dfrac{\Omega}{2}\widehat{H^2}(k) \right) \right], \\[3mm]
     &\mathbf{Y} + \dfrac{1}{2J}\left(\mathbf{\Phi}_\xi^2 - \mathbf{\Psi}_\xi^2 \right) +  \dfrac{1}{J}(F+\Omega \mathbf{Y})\mathbf{\Phi}_\xi X_\xi - \Omega \mathbf{\Psi}  + P(\mathbf{X}) = 0. \label{st-eqlast}
\end{empheq}
It is worth mentioning that since the system \eqref{st-eq1}-\eqref{st-eqlast} does not change on time, the unknowns do not depend on the variable $t$.

\subsection{Flow structure beneath the free surface wave}

In this subsection, we will show how to compute the streamlines, the stagnation points, and the pressure using the formulation from the previous subsections. In the stationary regime, particle pathlines coincide with streamlines. Therefore, given a particle that is initially located at the point $(x_0,y_0)$, its trajectory can be described by the following ODE system
\begin{equation*}
    \begin{cases}
        \dfrac{dx}{dt} = \phi_x(x,y) + \Omega y + F, \\[2mm]
        \dfrac{dy}{dt} = \phi_y(x,y), \\
        (x(0),y(0)) = (x_0,y_0).
    \end{cases}
\end{equation*}
Since $\psi$ is the harmonic conjugated of $\phi$, we can rewrite the previous system as
\begin{equation*}
    \begin{cases}
        \dfrac{dx}{dt} = \psi_y(x,y) + \Omega y + F =: \dfrac{\partial \psi_s}{\partial y}, \\[2mm]
        \dfrac{dy}{dt} = -\psi_x(x,y) =: -\dfrac{\partial \psi_s}{\partial x}, \\
        (x(0),y(0)) = (x_0,y_0).
    \end{cases}
\end{equation*}
Solving the system for $\psi_s$, we find that unless a constant
\begin{equation*}
     \psi_s(x,y) = \psi(x,y) + \dfrac{\Omega}{2}y^2 + Fy,
\end{equation*}
is the stream function. Denoting by $\overline{\psi}_s$ the stream function in the canonical variables, we have
\begin{equation}
     \overline{\psi}_s(\xi,\eta) = \overline{\psi}(\xi,\eta) + \dfrac{\Omega}{2}y(\xi,\eta)^2 + Fy(\xi,\eta), \label{stream-function-canonical-variables}
\end{equation}
where $\overline{\psi}$ and $y$  are given by equations \eqref{psi-cd} and \eqref{y-cd}.

The stagnation points are points that satisfy
\begin{equation*}
    \begin{cases}
        \psi_y(x,y) + \Omega y + F = 0, \\
        \psi_x(x,y) = 0.
    \end{cases}
\end{equation*}

We recall that $\overline{\psi}(\xi,\eta) = \psi(x(\xi,\eta),y(\xi,\eta))$, then if we derive in respect to the variables $\xi$ and $\eta$, we obtain that
 \begin{equation*}
     \begin{cases}
        \overline{\psi}_\xi = x_\xi \psi_x +  y_\xi \psi_y, \\
         \overline{\psi}_\eta =  x_\eta \psi_x +  y_\eta \psi_y.
     \end{cases}
 \end{equation*}
Solving the system of equations for $\psi_x$ and $\psi_y$, we find that
\begin{align*}
    &\psi_x = \dfrac{1}{|J|}(\overline{\psi}_\xi x_\xi - \overline{\psi}_\eta y_\xi), \\
    &\psi_y = \dfrac{1}{|J|}(\overline{\psi}_\eta x_\xi + \overline{\psi}_\xi y_\xi),
\end{align*}
where $|J|(\xi,\eta) := x_\xi^2 + y_\xi^2$. Hence, stagnation points can be found in the canonical domain by solving the system
\begin{align}
 & \dfrac{1}{|J|}(\overline{\psi}_\eta x_\xi + \overline{\psi}_\xi y_\xi)+ \Omega y(\xi,\eta) + F = 0\label{stag-points-canonical-domain1} \\
    &\dfrac{1}{|J|}(\overline{\psi}_\xi x_\xi - \overline{\psi}_\eta y_\xi) = 0.  \label{stag-points-canonical-domain2} 
   \end{align}

Lastly, to find the pressure in the fluid body we use the formula \eqref{bernoulli} and the Cauchy-Riemann equations which yields

\begin{equation*}
    p(x,y) = -\left(  \dfrac{1}{2}(\psi_x^2 +\psi_y^2) + (F+\Omega y)\psi_y + y-\Omega \psi \right).
\end{equation*}
Let $\overline{p}(\xi,\eta) = p(x(\xi,\eta),y(\xi,\eta))$. Thus, we obtain that
\begin{align}
        \overline{p}(\xi,\eta) &= -\bigg(  \dfrac{1}{2|J|^2}\left((\overline{\psi}_\xi x_\xi - \overline{\psi}_\eta y_\xi)^2 +\left(\overline{\psi}_\eta x_\xi + \overline{\psi}_\xi y_\xi\right)^2\right) \notag \\
        &\hspace{1.2cm}+ \dfrac{1}{|J|}(F+\Omega y(\xi,\eta))(\overline{\psi}_\eta x_\xi + \overline{\psi}_\xi y_\xi) + y(\xi,\eta)-\Omega \psi \bigg). \label{pressure-cd}
\end{align}

\section{Numerical scheme}\label{ns}

We notice that in the system \eqref{st-eq1}-\eqref{st-eqlast}, once $\mathbf{Y}$ is found, the other unknowns can be immediately computed. However, since we will need to know $H(\xi)$ beforehand, we first need a method to obtain it. Because of that, we use the following iterative scheme \cite{Flamarion:2019}:
\begin{empheq}[left=\empheqlbrace]{align}
        &H^m(\xi) = h(X_b^{m-1}(\xi)), \label{itm1} \\
        &X_b^{m+1}(\xi) = \xi + \mathcal{F}_{k \neq 0}^{-1}[i\tanh(kD)\hat{H}^m(k)] + \mathcal{F}_{k\neq 0}^{-1}\left[i\coth(kD)\left(\dfrac{\hat{H}^m(k)}{\cosh^2(kD)}-\dfrac{\hat{\mathbf{Y}}(k)}{\cosh(kD)}\right)\right], \label{itm2}
\end{empheq}
where the initial step is $X_b^0(\xi) = \xi$ and the stopping criteria is
\begin{equation*}
    |H^m(\xi) - H^{m-1}(\xi)| < \varepsilon,
\end{equation*}
where $\varepsilon > 0$ is a given tolerance.

Now, consider a discretization on a uniform grid in the variable $\xi$
\begin{equation*}
    \xi_j = -L + j\Delta \xi, j = 0,1,\dots,N-1, \Delta \xi = \dfrac{2L}{N},
\end{equation*}
and denote the functions evaluated on the grid points by $Y_j = \mathbf{Y}(\xi_j), X_j = \mathbf{X}(\xi_j), \Psi_j = \mathbf{\Psi}(\xi_j), \Phi_j = \mathbf{\Phi}(\xi_j), H_j = H(\xi_j)$ and $P_j = P(X_j)$. Thus, we have the following system of equations
\begin{equation}
    G_j(Y_0,Y_1,\dots,Y_{N-1}) = Y_j + \dfrac{1}{2J}\left(\Phi_{j,\xi}^2 - \Psi_{j,\xi}^2 \right) +  \dfrac{1}{J}(F+\Omega Y_n)\Phi_{j,\xi} X_{j,\xi} - \Omega \Psi_j  + P_j = 0, \quad j = 0,1,\dots,N-1.
\label{stsystem}
\end{equation}
The system \eqref{stsystem} is solved by using Newton's method. The Jacobian for Newton's method is computed using
\begin{equation*}
    \dfrac{\partial G_j}{\partial Y_l} = \dfrac{G_j(Y_0,Y_1,\dots,Y_l + \Delta Y,\dots,Y_{N-1}) - G_j(Y_0,Y_1,\dots,Y_l,\dots,Y_{N-1})}{\Delta Y}.
\end{equation*}
The stopping criteria is
\begin{equation*}
    \max_{0 \leq j \leq N-1} |G_j(Y_0,Y_1,\dots,Y_{N-1})| < \varepsilon.
\end{equation*}
where $\varepsilon > 0$ is a given tolerance. 
The solution is continued in the amplitude of $h$ by starting with a smaller amplitude than the desired. After computing $H$ through the scheme \eqref{itm1}-\eqref{itm2}, we use it to obtain the new $\mathbf{Y}$ and use the prior convergent solution as initial guess.

The stream function and pressure can be evaluated by formulas \eqref{stream-function-canonical-variables} and \eqref{pressure-cd}, in this order, which maps a uniform grid in the canonical domain onto a grid in the physical domain. These values are used to drawn level curves of $\overline{\psi}_s$ and $\overline{p}$. The stagnation points are computed by numerically solving the equation \eqref{stag-points-canonical-domain1} and \eqref{stag-points-canonical-domain2}. We observe that the required derivatives can be easily computed using elementary calculus. Once they are found in the canonical domain, we use the conformal map formulas \eqref{y-cd}-\eqref{x} to plot it in the physical domain.

\section{Results and Discussions}

In this section, we investigate numerically the structure of the fluid beneath stationary waves.
Namely, we analyze the influence of the vorticity in the emergence and disappearance of stagnation points and the pressure within the bulk of the fluid. It is worth to mention that in all of our experiments we have fixed the computational domain as $[-L,L) = [-100,100)$ with $N = 2^{10}$ uniformly spaced points, considered in the physical domain the bottom obstacle of the form 
\begin{equation}
    h(x) = Ae^{-0.1x^2}, \notag 
\end{equation}
where $A$ is the amplitude and set the pressure on the free surface as $P(x) = 0$. The case of a pressure distribution along the free surface will be the topic of another paper. We note that depending on the signal of $A$ we can have a bump ($A > 0$) or a hole ($A < 0$), for that reason we discuss each case separately.

We remark that in the absence of the vorticity the  fundamental parameter used for describing the pattern of waves generated due to a current–topography interaction is the Froude number. In this case the Froude number is critical when $F = 1$, i.e. when the linear long-wave phase speed
is equal to the mean flow speed. However in the presence of vorticity it shifts to \cite{Flamarion:2019} 
\begin{equation}
    F_c = -\dfrac{\Omega}{2} + \dfrac{\sqrt{\Omega^2 + 4}}{2}. \label{critical-froud-number}
\end{equation}
Moreover, we will study both subcritical (\(F < F_c\)) and supercritical (\(F > F_c\)) cases. The background shear flow depends on both \(\Omega\) and \(F\) (see equation \eqref{eq:shearflow}). By fixing \(F = 1.5\), we can transition between subcritical and supercritical flows by varying \(\Omega\). In this scenario, the flow becomes supercritical for \(\Omega > \Omega_0\) and remains subcritical for \(\Omega < \Omega_0\), where \(\Omega_0 =  -\frac{5}{6}\).

\begin{figure}[!htb]
    \centering
    \includegraphics[scale=0.7]{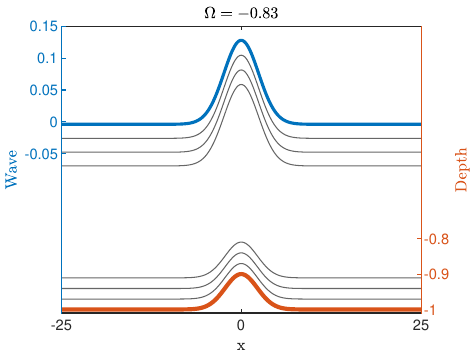}
    \includegraphics[scale=0.7]{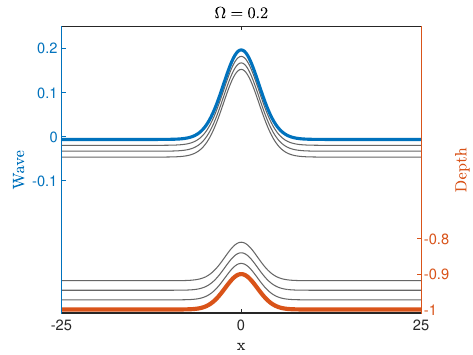}
    \includegraphics[scale=0.7]{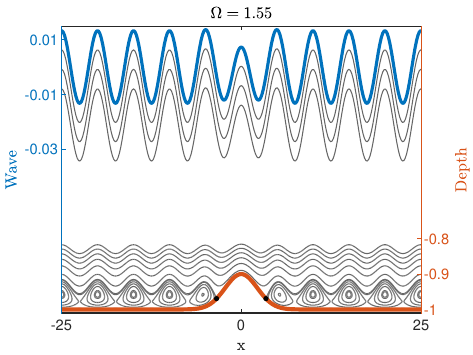}
    \includegraphics[scale=0.7]{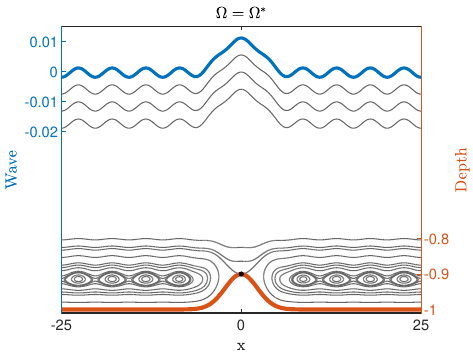}
    \includegraphics[scale=0.7]{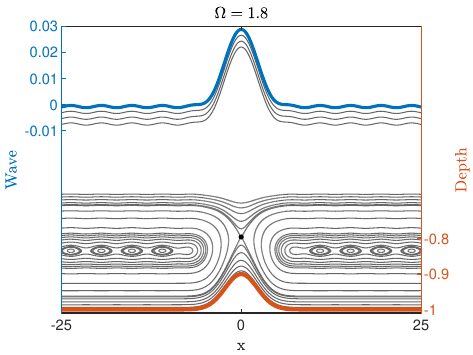}
    \includegraphics[scale=0.7]{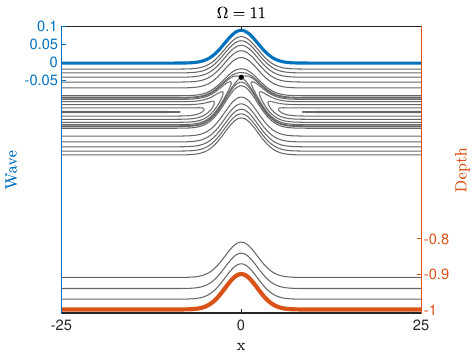}
     \caption{Phase portraits for waves in the supercritical regime. The left vertical axis represents the scale for the free surface disturbance, while the right vertical axis shows the depth scale, indicating the streamlines near the bottom.}
    \label{pp-supercritical-bump}
\end{figure}

\subsection{Streamlines and stagnation points for a bump}

We begin by investigating scenarios where the bottom obstacle profile is a bump. In this case we consider the topography height in the physical domain to be $A = 0.1$. In our first experiment we varied $\Omega$ while keeping it in the supercritical regime ($\Omega > \Omega_0)$. The resulting phase portraits can be seen in figure \eqref{pp-supercritical-bump} in which the stagnation points are shown with black circles.

Initially, we see that there are no stagnation points. However, as the value of $\Omega$ increases, they emerge from the bottom, resulting in the appearance of the cat's eye structure. As $\Omega$ increases further, two stagnation points, on the bottom, gradually get closer and merge into one at $x = 0$ for $\Omega^\star \approx 1.6401$ and the recirculation zones disattach from the lower boundary. For larger values of the vorticity, this stagnation point moves upward in direction to the free surface while the cat's eye structures shrink and disappear.

It is interesting to discuss about the particular stagnation point that remains when the vorticity $\Omega$ is large. Inspecting the portrait phases in figure \ref{pp-supercritical-bump}, we note that we have found a saddle point beneath the crest of the free surface wave. This is a novelty, since in the flat bottom case only centers points are found under the wave crest.

\begin{figure}[!htb]
    \centering
    \includegraphics[scale=0.8]{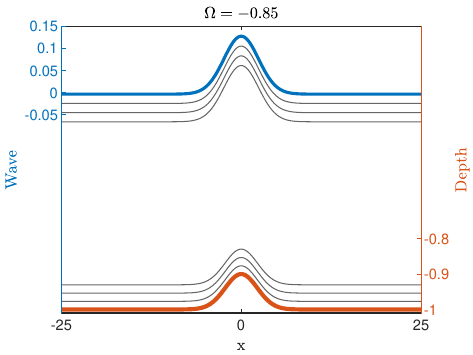}
    \caption{Phase portrait for waves in the subcritical regime. The left vertical axis represents the scale for the free surface disturbance, while the right vertical axis shows the depth scale, indicating the streamlines near the bottom.}
    \label{pp-subcritical-bump}
\end{figure}

Next, we focused in the subcritical regime, that is, we have varied the vorticity such that we had $\Omega < \Omega_0$. In this context, there are no stagnation points, therefore the cat's eyes structure does not appear in this regime. In figure \ref{pp-subcritical-bump}, we displayed a typical phase portrait.

\begin{figure}[!htb]
    \centering
    \includegraphics{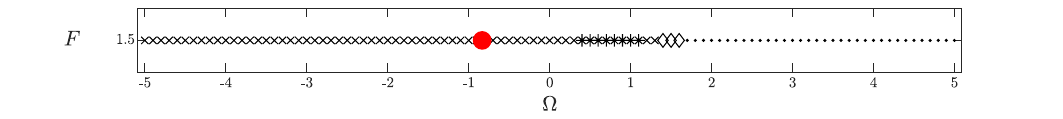}
    \caption{ 
     Dependence of stagnation point existence as vorticity ($\Omega$) varies, with $F$ held constant at $1.5$.
The symbol $\times$ represents values of vorticity where no stagnation points are found. The red circle marks the transition from subcritical to supercritical regimes. The symbol $\cdot$ corresponds to values of vorticity where stagnation points are located inside the fluid. The symbol $\diamond$ indicates values of vorticity where stagnation points are found both inside the fluid and on the bottom topography. The symbol $\ast$ represents values for which a stationary solution could not be obtained.
    }
    \label{bump-patterns}
\end{figure}

In Figure \ref{bump-patterns}, we illustrate the dependence of stagnation point existence as \( \Omega \) varies from \(-20\) to \(20\). For each chosen value of \( \Omega \), we computed the stationary solution (when feasible) and analyzed the flow to identify the presence of stagnation points.
The red circle indicates the transition from  subcritical to supercritical  regimes.  The dots marked with \( \times \) represent values of \( \Omega \) where no stagnation points are present. The dots marked with \( \cdot \) correspond to values of \( \Omega \) where stagnation points are located inside the fluid. Similarly, the dots marked with \( \diamond \) indicate values of \( \Omega \) where stagnation points are found both inside the fluid and on the topographical obstacle. Finally, the dots marked with \( \ast \) denote values for which a stationary solution could not be obtained. It is important do mention that since for $|\Omega| > 5$ the pattern remains the same, thus we restricted the visualization of the map to the interval $[-5,5]$.

 We have decided to classify the location of the stagnation points rather the quantity because as we could see in figure \ref{pp-supercritical-bump}, it is possible to have many stagnation points. This finding is intriguing since in the flat bottom scenario, this quantity varies between zero to three \cite{Ribeiro:2017}.

\subsection{Streamlines and Stagnation points for a hole}

\begin{figure}[!htb]
    \centering
    \includegraphics[scale=0.7]{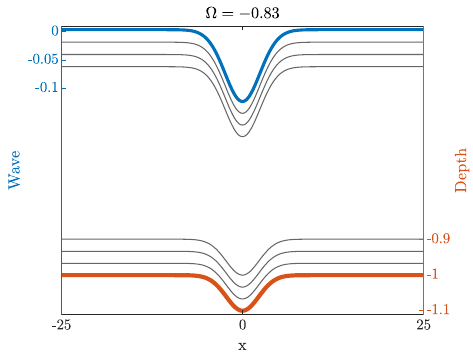}
    \includegraphics[scale=0.7]{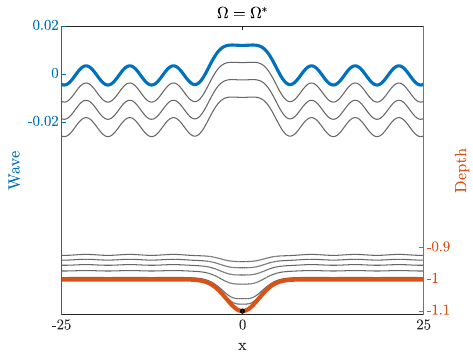}
    \includegraphics[scale=0.7]{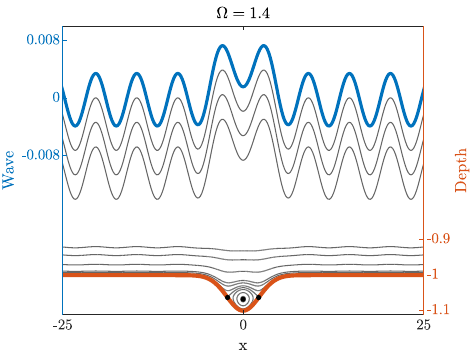}
    \includegraphics[scale=0.7]{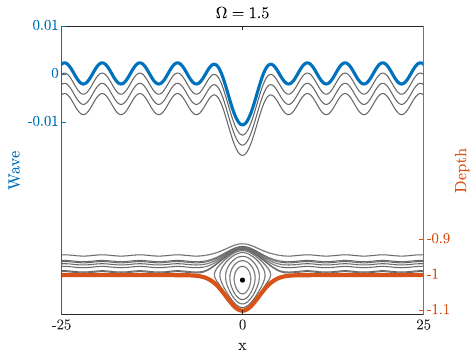}
    \includegraphics[scale=0.7]{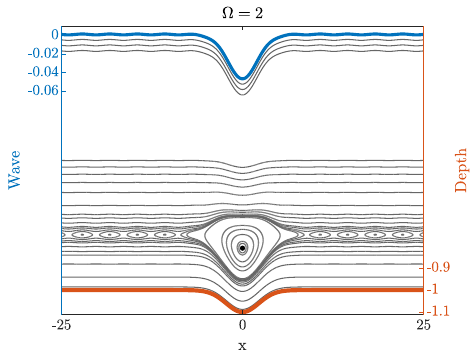}
    \includegraphics[scale=0.7]{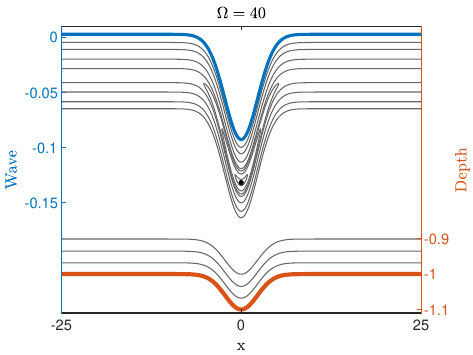}
\caption{Phase portraits for waves in the supercritical regime. The left vertical axis represents the scale for the free surface disturbance, while the right vertical axis shows the depth scale, indicating the streamlines near the bottom.}
     \label{hole-supercritical-case}
\end{figure}

Now we move to explore scenarios where the bottom obstacle profile is a hole, that is, when $A < 0$. Here, for our numerical simulations, we have fixed $A = -0.1$. In a similar fashion, we begin by investigating the supercritical case. We have found that there are no stagnation points for $\Omega < \Omega^\ast \approx 1.3414$. At $\Omega = \Omega^\ast$ a single stagnation point emerge on the bottom obstacle at $x = 0$. As we increase the vorticity, it splits in three points: two saddles (attached to the bottom) and one center (inside the fluid domain). Increasing it further results in stagnation points only inside the fluid and the appearance of the cat's eye structure. However, for large values of $\Omega$ this structure vanishes and the only stagnation point moves upward in direction of the surface wave. Those results can be observed in figure \ref{hole-supercritical-case}.

We highlight the novel finding of a stagnation point in the bulk of the fluid beneath a depression solitary wave. To the best of our knowledge, this is the first instance in the literature where such a phenomenon has been observed in the context of gravity waves. However, when surface tension or an electric field is taken into account, critical points can emerge in the fluid, as demonstrated in \cite{FlamarionGaoRibeiro:2023,Flamarion2:2023}.

\begin{figure}[!htb]
    \centering
\includegraphics[scale=0.7]{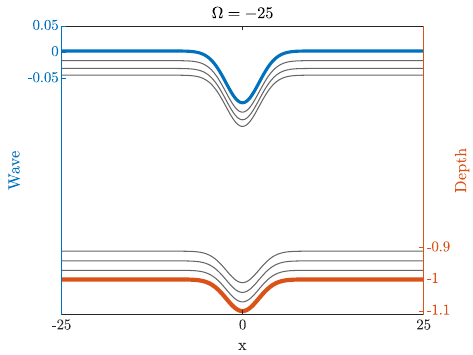}
    \caption{Typical phase portrait for waves in the subcritical regime. The left vertical axis represents the scale for the free surface disturbance, while the right vertical axis shows the depth scale, indicating the streamlines near the bottom.}
    \label{hole-subcritical-case}
\end{figure}

We now proceed to the subcritical case. As in the previous case, where the topography profile was a bump, no stagnation points were found. In figure \ref{hole-subcritical-case} we present a typical portrait phase for this scenario.

\begin{figure}[!htb]
    \centering
    \includegraphics{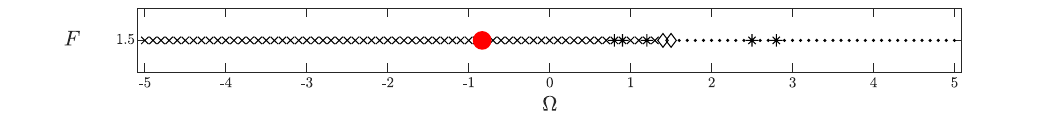}
    \caption{Dependence of stagnation point existence as vorticity ($\Omega$) varies, with $F$ held constant at $1.5$.
The symbol $\times$ represents values of vorticity where no stagnation points are found. The red circle marks the transition from subcritical to supercritical regimes. The symbol $\cdot$ corresponds to values of vorticity where stagnation points are located inside the fluid. The symbol $\diamond$ indicates values of vorticity where stagnation points are found both inside the fluid and on the bottom topography. The symbol $\ast$ represents values for which a stationary solution could not be obtained.}
    \label{hole-patterns}
\end{figure}

In Figure \ref{hole-patterns}, we present a graph illustrating the presence of stagnation points as \( \Omega \) varies, similar to the analysis in the previous subsection. Once again, we observe that there are no stagnation points in the subcritical regime.

\subsection{Pressure}

\begin{figure}[!htb]
\centering
\includegraphics{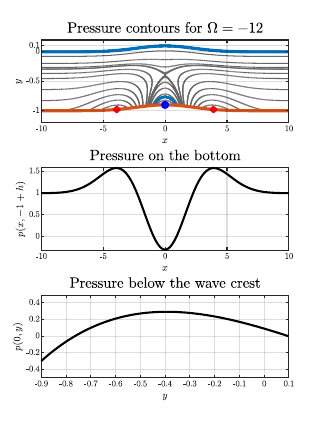}
\includegraphics{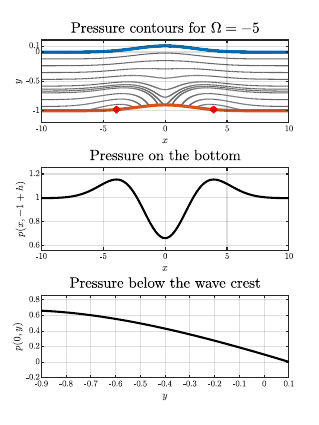}
\includegraphics{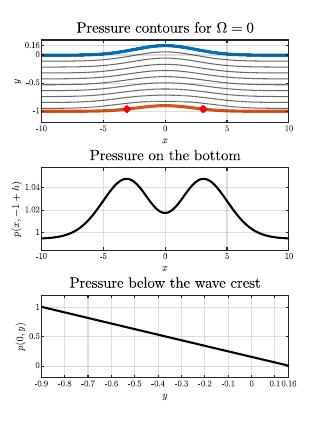}
\includegraphics{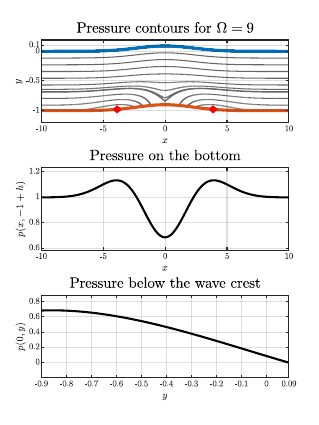}
\includegraphics{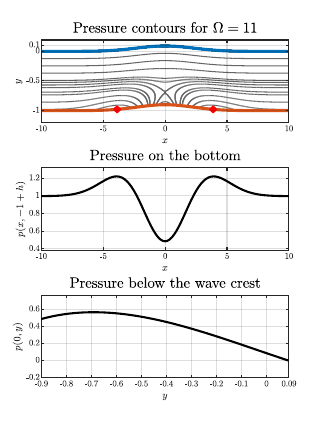}
\includegraphics{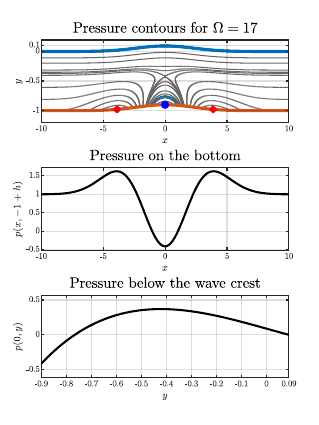}
    \caption{Pressure  beneath a wave over a bump bottom topography as vorticity is varied. Red dots indicate the locations of global pressure maxima, while blue dots represent the locations of global pressure minima. The blue line in the fluid body for the pressure contour $\Omega = -12$ and $\Omega = 17$ indicate points in which the pressure is equal to the atmospheric pressure. }
     \label{pressure-contour-bump}
\end{figure}

Now, we will examine the pressure within the bulk of the fluid domain. It is well known that for pressure beneath periodic waves in  irrotational flows \cite{ConstantinStrauss:2010}:
\begin{itemize}
    \item[(i)] the maximum of the pressure value is achieved at the point under the wave crest on the flat bed, and the minimum on the free surface;    
    \item[(ii)] the pressure strictly increases with the depth along vertical lines, and strictly decreases along a horizontal line starting below the crest and ending at the through.
\end{itemize}
In the context of flows with a constant vorticity, these results are no longer generally valid. Strauss and Wheller~\cite{StraussWheeler:2016} have rigorously proven that under certain conditions, for any wave (periodic or solitary), the pressure minimum value is achieved inside the fluid instead on the free surface. Ribeiro-Jr {\it et al.} . \cite{Ribeiro:2017} have shown numerically that, for periodic waves, when the vorticity is sufficiently strong, the global minimum pressure shifts from the free surface to a point beneath the crest. Similarly, Flamarion {\it et al.} . \cite{Castro:2023} reported analogous results for solitary waves. Moreover, it is known that for periodic waves, the global maximum pressure can occur below the trough \cite{Ribeiro:2017}.

Figure \ref{pressure-contour-bump} displays the pressure contour, the pressure on the bottom boundary and below the wave crest for various values of $\Omega$ for the case in which we have a bump on the bottom topography. The blue dots represent the point of global minimum pressure, when it is below the pressure on the surface wave, and the red dots indicate the point of global maximum of pressure. As the absolute value of the vorticity increases, we observe the emergence of a region in which the pressure is lower than the atmospheric and parameter regimes where the pressure  reaches its minimum on the bottom topography, as well as we see the pressure below the wave crest no longer being monotonic. 

\begin{figure}[!htb]
    \centering
\includegraphics{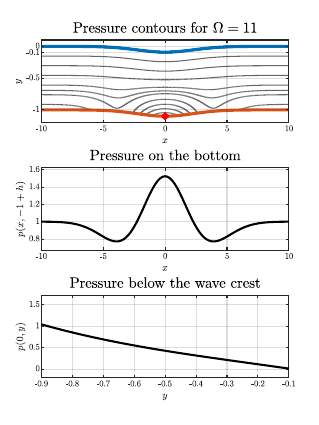}
\includegraphics{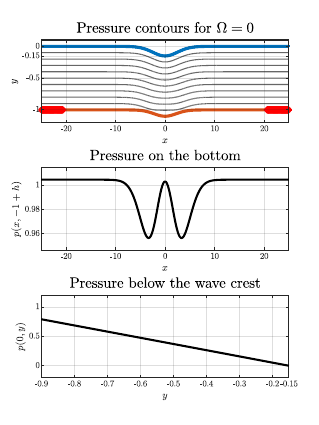}

\includegraphics{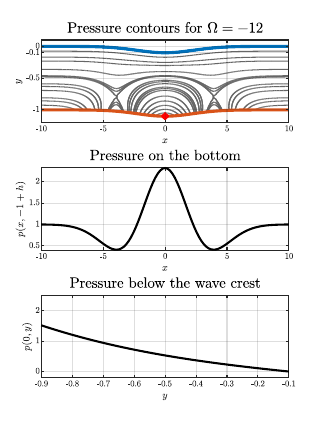}
\includegraphics{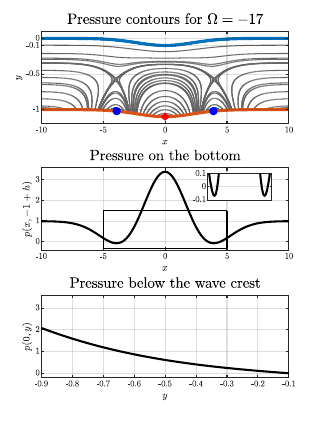}
    \caption{Pressure  beneath a wave over a hole bottom topography as vorticity is varied. Red and blue dots represent the localization of global maxima and minima of pressure respectively.}
    \label{pressure-contour-hole}
\end{figure}

We also computed the pressure in the presence of a bottom topography featuring a hole. In this scenario, as the vorticity decreases, the global minimum pressure shifts from the surface to the bottom boundary. To the best of our knowledge, this is the first time such a pressure anomaly has been reported in the literature. Notably, in this case, we observe the emergence of two points of global minimum pressure, in contrast to the previous scenario.These results can be seen in figure \ref{pressure-contour-hole}.

It is important to emphasize that pressure anomalies beneath gravity waves in flows with a flat bottom are typically associated with the presence of stagnation points, as indicated in previous studies \cite{Ribeiro:2017,Doak:2022, Castro:2023}. However, as shown in Figures \ref{bump-patterns} and \ref{hole-patterns}, no stagnation points are observed in the flow for $\Omega < \Omega_0 = -5/6$. Despite this, the flow exhibits pressure anomalies, such as a global minimum pressure on the bottom boundary, for values of $\Omega < \Omega_0$.

\section{Conclusion}

In this paper, we investigated the flow structure beneath stationary nonlinear surface waves influenced by bottom topography in a fluid with constant vorticity. We analyzed two types of topography: a bump and a hole. Our findings reveal that stagnation points occur exclusively in the supercritical regime, with their number varying between zero, one, three, or multiple points depending on the vorticity strength. Notably, when the topography featured a bump, a saddle point was observed beneath the wave crest. For both topography types, we found that increasing the absolute value of the vorticity causes the global minimum pressure to shift from the free surface to the bottom boundary. To the best of our knowledge, this is the first study to report such a pressure anomaly in the literature. Furthermore, we showed that pressure anomalies can arise even in the absence of stagnation points within the flow. In cases where the topography featured a hole, stagnation points were observed beneath depression solitary waves.

 \section*{Acknowledgments}
The research of the author L.G.M. was partially funded by the Coordenação de Aperfeiçoamento de Pessoal de Nível Superior – Brasil (CAPES) under Finance Code 001 (L.G.M.). The work of M.V.F. and R.R. Jr. was supported by the National Council for Scientific and Technological Development (CNPq) through the Call CNPq/MCTI Nº 10/2023 – Universal.

\end{document}